\documentclass[preprint,5p,twocolumn,10pt]{elsarticle}

\usepackage{bm,amsmath,amsthm,amssymb,amsfonts}
\usepackage{graphicx}
\usepackage{hyperref}
\DeclareMathOperator{\sgn}{sgn}

\journal{Computer Physics Communications}
\biboptions{sort&compress}

\begin{document}

\begin{frontmatter}

\title{First-principles calculation of higher-order elastic 
       constants from divided differences}

\author[a]{Ruvini Attanayake}
\author[a]{Umesh C. Roy}
\author[b]{Abhiyan Pandit}
\author[a,c]{Angelo Bongiorno\corref{cor}}

\cortext[cor]{Angelo Bongiorno. \textit{angelo.bongiorno@csi.cuny.edu}}

\address[a]{Department of Chemistry, College of Staten Island, Staten Island, NY 10314, USA}
\address[b]{Department of Chemistry and Biochemistry, California State University Northridge, CA 91330, USA}

\address[c]{The Graduate Center of the City University of New York, New York, NY 10016, USA}

\begin{abstract}
A method is presented to calculate from first principles 
the higher-order elastic constants of a solid material. 
The method relies on finite strain deformations, a density 
functional theory approach to calculate the Cauchy stress tensor, 
and a recursive numerical differentiation technique 
homologous to the divided differences polynomial 
interpolation algorithm. The method is applicable as is to any 
material, regardless its symmetry, to calculate elastic constants of, 
in principle, any order.
Here, we introduce conceptual framework and technical details 
of our method, we discuss sources of errors, we assess convergence 
trends, and we present selected applications. 
In particular, our method is used to calculate elastic 
constants up to the 6$^{th}$ order of two crystalline 
materials with the cubic symmetry, silicon and gold. 
To demonstrate general applicability, our method is also 
used to calculate the elastic constants up to the 5$^{th}$ 
order of $\alpha$-quartz, a crystalline material belonging 
to the trigonal crystal system, and the second- and 
third-order elastic constants of kevlar, a material 
with an anisotropic bonding network.
Higher order elastic constants computed with our method 
are validated against density functional theory calculations 
by comparing stress responses to large 
deformations derived within the continuum approximation.
\end{abstract}


\end{frontmatter}

\section{Introduction}\label{intro}

The elastic constants of a material encode in 
tensorial form the relationship between stress and 
strain \cite{c11,t24,tb64,w67}. 
The second-order elastic constants (SOECs) 
govern the linear terms of this relationship \cite{c11}, 
whereas third-, fourth-, and higher-order 
elastic constants carry information about 
elastic anharmonic properties \cite{br65,bf67,ccb18,bb22,l21}, 
the large deformation regime \cite{hg66,hk81,ps68,czl20}, 
and mechanical instabilities \cite{czl20,zcl18,bl20}. 
While SOECs have been determined for a large class of 
materials and can be easily calculated by using a density functional 
theory (DFT) approach \cite{jca15}, both measurement
and first-principles 
calculation of nonlinear elastic constants remain, in 
general, challenging tasks. 
As a result, third-order elastic constants (TOECs) 
are known for a restricted class of systems 
\cite{tb64,b65,tma66,pb83,t68,n71,lg11,tek17,jd06}, 
fourth-order elastic constants (FOECs) have been measured 
\cite{ps68,p69,g72,ksa97,ps03} and calculated \cite{wl09,wgz16,kv19,czl20,pb23cpc}
only for a handful of high-symmetry materials, 
whereas elastic constants of the 5$^{th}$ and higher order 
(HOECs) are practically absent from literature. 
General methods to calculate HOECs of, 
in principle, any order are needed to expand the 
capabilities of modern DFT-based strategies to predict 
materials properties, as well as to compensate for the 
lack of experimental information on nonlinear elastic 
behaviors of low-symmetry materials. 
Here, we present a potential solution to 
this problem.

SOECs are directly related to both sound velocity and 
macroscopic elastic moduli \cite{c11,w67}, 
and they carry information about 
the mechanical stability of a crystalline phase of 
a material \cite{wy93,wly95}.
TOECs govern the leading nonlinear terms of the stress-strain 
relationship \cite{c11,w67}, and they are intimately connected 
to thermal expansion \cite{rm73,wmp18,bb22}, sound propagation 
and attenuation \cite{tb64}, and the anharmonic character of 
the long-wavelength acoustic vibrational modes of a 
material \cite{br65,bf67}.
FOECs and HOECs are important to describe the stress 
response to large deformations \cite{cb67,g72,gg75}.
Furthermore, knowledge of these higher-order 
coefficients can give key insight into the onset, thermodynamics, 
and nature of solid phase transitions 
\cite{zcl18,bl20,czl20,l21,tt87,s82,tfa83,h94,teb17}.
%
%
Unlike SOECs, TOECs and especially FOECs and HOECs 
are difficult to measure \cite{tb64,ps68,g72,ksa97}.
TOECs are typically obtained from acousto-elastic 
experiments, by measuring the sound velocity in a material 
under different stress conditions \cite{tb64,kb65,hg66,tma66,jh67,t68,n71}.
Ultrasonic harmonic generation \cite{pb83} 
and shock wave compression \cite{lg11,g72} have also 
been used to measure these nonlinear elastic coefficients. 
Overall, over the past decades these experimental 
techniques have been used to determine the TOECs 
of various materials, predominantly crystalline solids 
belonging to the cubic and hexagonal systems 
\cite{tb64,ma64a,ma64,b65,pb83,t68,hs70,n71,lg11,tek17,jd06}, 
and isotropic materials \cite{b65,g72,sss93,wlq01}.
On the contrary, while experimental determination of 
FOECs has been attempted a 
handful of times \cite{kls81,kkk15,aqa22}, to the best of our 
knowledge, measurements of elastic constants of the 
5$^{th}$ and higher order remain yet to be reported.

Two are the techniques typically used 
to calculate the nonlinear elastic constants of 
a material by using a DFT approach \cite{nsg71,zmg07,wl09,vkl16,jwc17,tek17,lls21,ccb18,pb23prm}.
The traditional technique relies on fitting a 
dataset of energy or stress versus strain values 
via polynomial functions of the 
strain \cite{zmg07,wl09,vkl16,jwc17,lls21,czl20}. 
In this technique, the fitting procedure becomes 
increasingly cumbersome with both 
lowering the symmetry of the material and increasing 
the order of the elastic constants to be determined 
\cite{zmg07,wl09,vkl16,jwc17,lls21,czl20}.
As a result, and in spite of recent accomplishments \cite{czl20}, 
applications of this technique remain confined to the 
calculation of only TOECs of materials belonging to the
cubic or hexagonal crystal systems \cite{zmg07,wl09,vkl16,jwc17,lls21}.
An alternative technique to calculate 
TOECs relies on numerical differentiation and the use 
of finite difference formulas \cite{ccb18}.
In recent years, this approach has been applied to 
both 2D and 3D materials, belonging to the cubic, 
hexagonal, orthorhombic, and monoclinic crystal 
systems \cite{ccb18,bb22,pb23prm}. Furthermore, 
this technique has been recently extended to calculate 
FOECs of cubic and hexagonal crystalline 
materials \cite{pb23cpc}. This latest formulation 
employs two and three multivariate finite-difference formulas to 
calculate TOECs and FOECs, respectively \cite{ccb18,pb23cpc}.
Further extension of this technique (to lower-symmetry 
materials and/or elastic constants of the 5$^{th}$ or 
higher order) require the use of 
higher-order multivariate finite-difference formulas, 
resulting in a cumbersome numerical scheme 
yielding HOECs subjected to a nonuniform
range of truncation errors.

Inspired by latest developments \cite{ccb18,pb23cpc}, 
here we present a method to calculate HOECs based on 
the use of finite deformations and a recursive numerical 
differentiation 
technique homologous to the divided differences polynomial 
interpolation algorithm \cite{mt64}. 
In contrast to existing techniques, the present method 
is technically complete, meaning that it is applicable 
as is to any material, irrespective of its symmetry, 
to calculate elastic constants of, in principle, any order. 
Here, we first introduce and discuss conceptual and 
technical aspects of our method (Sec. \ref{meth}). 
Errors and workflow of the method, and technical specifications 
of the software are discussed in Sec. \ref{erro}, 
whereas technical details of DFT calculations are 
discussed in Sec. \ref{dft}.
Then, we use our method to calculate the elastic constants 
up to the 6$^{th}$ order of both silicon and gold (Sec. \ref{siau}). 
To demonstrate validity, 
besides comparisons with published data, 
the computed elastic constants are supplied to a 
nonlinear elastic constitutive scheme to predict, 
and then compare to DFT results, the elastic responses 
to large uniaxial, shear, and hydrostatic deformations. 
To demonstrate general 
applicability and show potential applications 
of our method, we first calculate elastic constants up to 
the 5$^{th}$ order of $\alpha$-quartz, and we exploit 
these coefficients to gain insight on its 
the pressure-induced amorphization transition (Sec. \ref{quar}).
Then, we calculate SOECs and TOECs of 
kevlar (Sec.\ \ref{kevlar}), a material exhibiting an 
anisotropic bonding 
network, held together by covalent, van der Waals, and 
hydrogen-bond interactions.

\section{Method}\label{meth}

\subsection{Background and method formulation}

Our method rests on finite strain theory \cite{c11,w67}.
For completeness, here we introduce key formulas and notation. 
The Helmholtz free energy per unit mass, $A$, of a 
solid material can be written, upon selection of 
a reference state, as a Taylor series 
expansion in terms of the Green-Lagrangian strain \cite{c11,w67}:
\begin{eqnarray} \label{interu}
	\rho_0 A(\vec{\mu}) = \rho_0 A_0 &+& 
       \frac{1}{2} C^{(2)}_{\alpha_{1} \alpha_{2}}
       \mu_{\alpha_{1}} \mu_{\alpha_{2}} + \nonumber \\ 
       && \frac{1}{6} C^{(3)}_{\alpha_{1} \alpha_{2} \alpha_{3}}
   \mu_{\alpha_{1}} \mu_{\alpha_{2}} \mu_{\alpha_{3}} + \cdots + \nonumber \\
 && \frac{1}{n!} C^{(n)}_{\alpha_{1} \cdots \alpha_{n}}
     \mu_{\alpha_{1}} \cdots \mu_{\alpha_{n}} + \cdots,
\end{eqnarray}
where $\rho_0$ is the mass density of the material in the 
reference state, $A_0$ is the corresponding free energy per 
unit mass, $\mu_{\alpha_{i}}$ are components of the 
strain tensor $\bm{\mu}$, $\alpha_{i}$ are Voigt indices 
(ranging from 1 to 6, such that 
1 $\rightarrow$ xx, 2 $\rightarrow$ yy,
3 $\rightarrow$ zz, 4 $\rightarrow$ yz, 5 $\rightarrow$ zx,
and 6 $\rightarrow$ xy), and
$C^{(n)}_{\alpha_{1} \cdots \alpha_{n}}$ are 
the (isothermal) elastic constants of $n^{th}$-order.
From Eq.\ \ref{interu}, we can write:
\begin{eqnarray}\label{noecs}
C^{(n)}_{\alpha_{1} \alpha_{2} \cdots \alpha_{n}} &=& 
\rho_0 \left.  \frac{\partial^n A} {\partial \mu_{\alpha_{1}} 
\partial \mu_{\alpha_{2}} \cdots \partial \mu_{\alpha_{n}} }
\right|_{\bm{0}}  \nonumber \\
 &=& \left. \frac{\partial^{n-1} P_{\alpha_{1}} (\vec{\mu})}
{\partial \mu_{\alpha_{2}} \cdots \partial \mu_{\alpha_{n}}}
\right|_{\bm{0}},
\end{eqnarray}
where derivatives are taken at constant (temperature and) 
volume, i.e $\vec{\mu}=\bm{0}$, and $\bm{P}$ is the 
second Piola-Kirchhoff (PK2) stress 
tensor \cite{c11,w67,ccb18}, which satisfies the 
following relationships \cite{c11,w67,ccb18}:
\begin{eqnarray}\label{nle}
P_{\alpha_{1}} &=& \left.
\rho_0 \frac{\partial A}{\partial \mu_{\alpha_{1}}} \right|_{\bm{0}}  \nonumber \\
	\bm{P}(\bm{\mu}) &=& \det{[\bm{F}]} 
	\left( \bm{F}^{-1} \bm{\sigma} \bm{F}^{-T} \right) \nonumber \\
\bm{\mu} &=& \frac{1}{2} (\bm{F}^T \bm{F} - \bm{I}),
\end{eqnarray}
where 
$\bm{F}$ is the deformation gradient, and 
$\bm{\sigma}$ and $\bm{P}$ are Cauchy and PK2 stress 
tensors of the material subjected the strain 
$\bm{\mu}$, respectively. From Eqs. \ref{interu} 
and \ref{nle}, it trivially follows that:
\begin{equation} \label{pk2nu}
\begin{split}
P_{\alpha_{1}}(\vec{\mu}) &= P^0_{\alpha_{1}} +
C^{(2)}_{\alpha_{1} \alpha_{2}} \mu_{\alpha_{2}} +
\frac{1}{2} C^{(3)}_{\alpha_{1} \alpha_{2} \alpha_{3}}
\mu_{\alpha_{2}} \mu_{\alpha_{3}} \\
&+ \cdots + \frac{1}{(n-1)!} C^{(n)}_{\alpha_{2} \cdots \alpha_{n}}
        \mu_{\alpha_{2}} \cdots \mu_{\alpha_{n}} + \cdots,
\end{split}
\end{equation}
where $P^0_{\alpha_{1}}$ is equal to 
$\sigma^0_{\alpha_{1}}$, i.e. the Cauchy stress tensor 
of the material in the reference state.

Our method has its foundation in the 
following {\it ad hoc} generalization of Eq.\ \ref{noecs}:
\begin{equation}\label{gencn}
\mathfrak{C}^{(n)}_{\alpha_{1} \cdots \alpha_{n}} (\vec{\mu}^{\prime}) =
\left. \frac{\partial^{n-1} P_{\alpha_{1}} (\vec{\mu})} 
{\partial \mu_{\alpha_{2}} \cdots \partial \mu_{\alpha_{n}}}
\right|_{\vec{\mu}^{\prime}},
\end{equation}
defining a n$^{th}$-order elastic constant 
of a material accommodating a finite 
strain $\vec{\mu}^{\prime}$. Although physically 
meaningful only when $\vec{\mu}^{\prime}=\bm{0}$, 
Eq.\ \ref{gencn} is mathematically sound, 
and as such it can be exploited as follows. 
First, we rewrite Eq.\ \ref{gencn} as:
\begin{eqnarray}\label{gencnrec}
\mathfrak{C}^{(n)}_{\alpha_{1} \cdots \alpha_{n}} (\vec{\mu}^{\prime}) &=&
\left. \frac{\partial
\mathfrak{C}^{(n-1)}_{\alpha_{1} \cdots \alpha_{n-1}} (\vec{\mu})}
{\partial \mu_{\alpha_{n}}}\right|_{\vec{\mu}'} \nonumber \\
&=& \left. \frac{\partial^2
\mathfrak{C}^{(n-2)}_{\alpha_{1} \cdots \alpha_{n-2}}
        (\vec{\mu})}{\partial \mu_{\alpha_{n-1}}
                     \partial \mu_{\alpha_{n}}
        }\right|_{\vec{\mu}'} = \cdots \nonumber \\
&=& \left. \frac{\partial^{n-1} \mathfrak{C}^{(1)}_{\alpha_{1}} (\vec{\mu})}
{\partial \mu_{\alpha_{2}} \cdots \partial \mu_{\alpha_{n}}}
\right|_{\vec{\mu}^{\prime}},
\end{eqnarray}
where, for notational convenience,
$\mathfrak{C}^{(1)}_{\alpha_{1}}$ is used to refer to
the PK2 stress components $P_{\alpha_{1}}$.
Then, we express Eq.\ \ref{gencnrec} in a numerically 
convenient form in terms of a first-order finite 
difference operator, $\mathbb{D}_{\alpha}$, as follows:
\begin{eqnarray}\label{opera}
\mathfrak{C}^{(n)}_{\alpha_{1} \cdots \alpha_{n}} (\vec{\mu}^{\prime}) &=&
\mathbb{D}_{\alpha_{n}}
\mathfrak{C}^{(n-1)}_{\alpha_{1} \cdots \alpha_{n-1}} (\vec{\mu}^{\prime}) \nonumber \\
&=& \mathbb{D}_{\alpha_{n}} \mathbb{D}_{\alpha_{n-1}} 
\mathfrak{C}^{(n-2)}_{\alpha_{1} \cdots \alpha_{n-2} } (\vec{\mu}^{\prime}) 
= \cdots \nonumber \\
&=& \mathbb{D}_{\alpha_{n}} \mathbb{D}_{\alpha_{n-1}} \ldots 
\mathbb{D}_{\alpha_{2}} \mathfrak{C}^{(1)}_{\alpha_{1}}.
\end{eqnarray}
In particular, to the lowest order of accuracy, 
$\mathbb{D}_{\alpha}$ can be taken to be 
the following central first-order finite difference operator:
\begin{equation}\label{cfd2a}
\mathbb{D}_{\alpha} \mathcal{C} (\vec{\mu}) =
\frac{\mathcal{C} (\vec{\mu}+\xi \vec{1}_{\alpha} ) -
   \mathcal{C} (\vec{\mu}-\xi \vec{1}_{\alpha} )} { 2 \xi} ,
\end{equation}
where $\xi$ is a (constant and small) strain parameter, and 
$\vec{1}_{\alpha}$ is a six-dimensional strain (Voigt) 
vector with only one component $\alpha$ equal to 1 
and all the others equal to zero. 
Equations \ref{opera} and \ref{cfd2a} are the key formulas 
underlying the present method to calculate HOECs.

\begin{figure}[ht!]
\begin{center}
\includegraphics[width=\columnwidth]{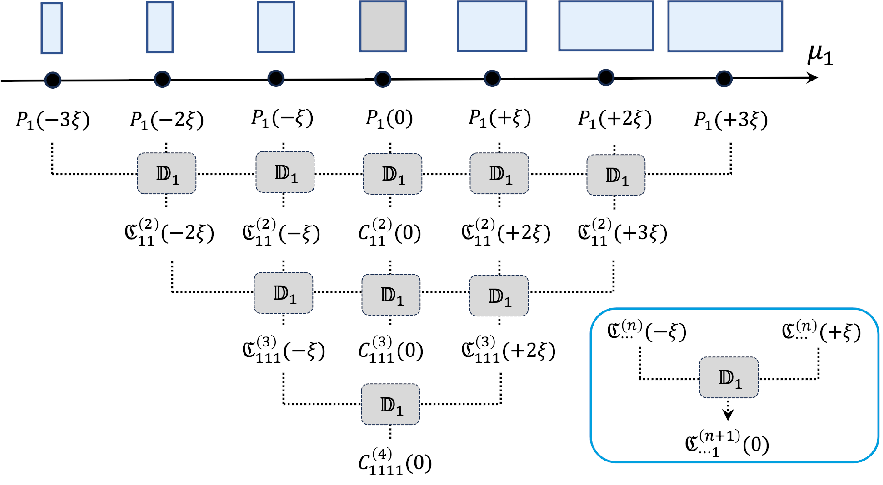}
\caption{\label{scheme} Schematic diagram illustrating 
the numerical basis of our method to 
calculate HOECs, in the specific case of elastic 
constants with equal indices (equal to $1\rightarrow xx$) 
up to the 4$^{th}$. From top to bottom, the PK2 stress 
tensor is sampled at equispaced finite strains around 
the reference state (gray square). Then, the 
first-order finite difference operator in Eq.\ \ref{cfd2a}
(inset, bottom right) is applied recursively (Eq.\ \ref{opera}), 
first to calculate elastic constants of 2$^{nd}$ order 
for reference and deformed states (light blue 
rectangles), and then to calculate derivatives of 
increasing order on a reducing number of deformed states, 
until the elastic constant of 4$^{th}$ order is calculated 
for only the reference state. }
\end{center}
\end{figure}

In essence, our method consists of a
numerical technique to calculate higher-order multivariate 
derivatives (Eqs.\ \ref{opera} and \ref{cfd2a}), inspired 
to the divided differences polynomial interpolation 
algorithm \cite{mt64} (Fig.\ \ref{scheme}).
As such, our method yields all the components of 
the elastic tensors up to a certain order that 
automatically possess the expected Voigt (index permutation) 
symmetry \cite{t24}.
Rotational symmetries leaving invariant the structure of the 
reference state are also transferred to the tensors through the DFT 
calculations. Numerical errors stemming from the DFT 
calculations introduce
small differences in elastic constants that are expected to be
equivalent, dependent, or zero. These differences can be
reduced by adopting the following optional solutions. 
First, by imposing rotational symmetry (consistent with
the symmetry operations of the reference state)
on the stress tensors of the various deformed
configurations used to compute the elastic tensors ({\it vide infra}).
Besides reducing numerical errors, this operation can 
also allow to saving a considerable amount of computational time.
Second, by enforcing permutational
symmetry on the elastic tensors of increasing order 
that are calculated recursively via Eqs.\ \ref{opera} and \ref{cfd2a}.
Third, by carrying out the following symmetry-generated averages:
\begin{equation}\label{sym}
\overline{C}^{(n)}_{ij\ldots} =
\frac{1}{N} \sum_{\alpha=1}^N
Q^{(\alpha)}_{iI} Q^{(\alpha)}_{jJ} \ldots C^{(n)}_{IJ\ldots},
\end{equation}
where $\bm{Q}^{(\alpha)}$ are the 3$\times$3
rotation matrices representing the $N$ symmetry
operations leaving invariant the structure of the
reference state, and $C^{(n)}_{IJ\ldots}$ and
$\overline{C}^{(n)}_{ij\ldots}$ are
components of the computed and symmetrized
elastic tensors of order $n$, respectively,
both expressed in terms of cartesian indices \cite{t24}.
Elastic constants reported in this work have been
obtained by adopting the aforementioned three practices.

\begin{figure}[ht!]
\begin{center}
\includegraphics[width=\columnwidth]{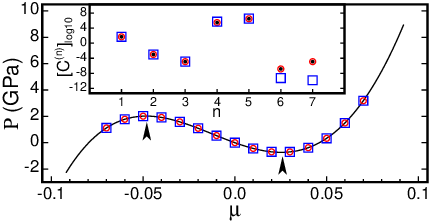}
\caption{\label{proof} Polynomial of
degree 10 (solid black line) reproducing a 
stress-strain curve, $P(\mu)$, encompassing
mechanical instabilities (black arrows).
Colored symbols show the data points
used to calculate derivatives of $P(\mu)$ at $\mu=0$ 
up to 7$^{th}$ order using Eqs.\ \ref{opera} and \ref{cfd2a}.
Exact values of $P(\mu)$ 
at equispaced strains are shown using red 
circles, whereas non-exact values including a 
random error of up to $\pm 0.01$ (GPa)
are shown using blue squares. Derivatives of $P(\mu)$ 
obtained by using the two sets 
of points are shown in the inset using the same
symbols, together with exact results (black discs). 
$[C^{(n)}]_{log10}$ stands for 
$\sgn{[C^{(n)}]} \times \log{|C^{(n)}|}$.}
\end{center}
\end{figure}

\subsection{Error analysis, workflow, and software}\label{erro}

As a proof of concept, to introduce technical 
aspects and discuss errors, we use a simple toy model.
In particular, we use an {\it ad hoc} polynomial function of 
degree 10 reproducing the stress versus strain 
curve, $P(\mu)$, shown in Fig. \ref{proof},
with polynomial coefficients, $C^n$, yielding 
instability points on both sides of the reference
state at zero pressure. 
To calculate these same 
coefficients using our method, we sample $P(\mu)$ 
at equispaced discrete points, and 
then we use Eqs. \ref{opera} and \ref{cfd2a} to 
calculate the derivatives of $P(\mu)$ at $\mu=0$.
This simple application shows that our method yields 
$C^n$ in excellent agreement with exact values, 
regardless of the strain spacing used to generate 
the sampling grid. In detail, strain parameters 
between 0.002 and 0.015 
yield values of $C^n$, $n=1 \ldots 10$, 
deviating 4\% at most from exact values. 
This result demonstrates that the present 
technique is sound and accurate, and that 
truncation errors introduced by the use of 
Eq.\ \ref{cfd2a} are not a major source of inaccuracy.

The principal source of errors is extrinsic to our 
method, and it lies in the numerical precision of the 
discrete values of $P(\mu)$, supplied to 
Eqs. \ref{opera} and \ref{cfd2a}. 
In fact, Fig. \ref{proof} shows that if sampling 
is not exact, i.e. if the discrete values of 
$P(\mu)$ include small random errors, then our 
method yields derivatives whose accuracy decreases 
for increasing the order (see caption of Fig. \ref{proof} 
for details). Overall, the results obtained by using 
this and other toy models consistently show that (i) 
our method is sound, (ii) truncation errors due to 
the use of Eq.\ \ref{cfd2a} have a small impact on 
overall accuracy, and most importantly (iii), 
the loss of accuracy in 
evaluating derivatives of increasing order 
arises from the lack of numerical precision 
of the stress values sampled at equispaced 
strains \cite{ah99}.

In practice, our method to calculate HOECs of a material 
treated at a DFT level entails the following workflow.
\begin{itemize}
\setlength{\itemindent}{0em}
\item
First, generation of a grid of deformed 
configurations centered around a reference 
state. In detail, when the multivariate 
derivatives in Eq.\ \ref{opera} are expressed 
in terms of the first-order finite-difference 
operator $\mathbb{D}_{\alpha}$ in Eq.\ \ref{cfd2a}, 
the grid of deformed configurations needed 
to calculate elastic constants 
up to the $n^{th}$ order is obtained 
by applying to the reference state the 
(Voigt) Lagrangian strain 
vectors $\vec{\mu} = \xi \left( i , j , k, p , q, r \right)$, 
where the integer components are such 
that $-n+1 \le i+j+k+p+q+r \le n-1$, and 
$\xi$ is a strain parameter, typically 
ranging between 0.005 and 0.02.

\item
Second, use of a DFT approach to optimize ionic 
positions and calculate the 
Cauchy stress tensor arising in each deformed 
configuration of the reference state. 
This task is computationally demanding. To reduce 
the number of calculations, we exploit, if present, 
the symmetry of the material. In particular, 
if $\bm{Q}$ is a symmetry 
operation leaving invariant the structure of 
a crystalline material in the selected 
reference state, we first identify 
the deformed configurations equivalent by 
symmetry, i.e. such that 
$\bm{Q}^T \bm{\mu}_a^{\prime} \bm{Q} = \bm{\mu}_b$, 
where $\bm{\mu}_a$ and $\bm{\mu}_b$ are 
strain matrices belonging to 
the grid of finite deformations introduced in the 
previous point. Then, we carry out a DFT calculation 
for only one representative member of the 
family, and we calculate the Cauchy stress tensor of 
the redundant states equivalent by symmetry as 
follows: $\bm{Q}^T \bm{\sigma}_a^{\prime} \bm{Q} = \bm{\sigma}_b$, 
where $\bm{\sigma}_a$ and $\bm{\sigma}_b$ refer to 
the stress tensors of the representative and 
redundant members, respectively.
It is to be noted that, thanks to the use of 
symmetry operations, the number of DFT calculations 
required to calculate the elastic constants up 
to the 6$^{th}$ order of, for example, the diamond 
phase of silicon reduces from 3653 to only 297.

\item
Third, given the full list of strain vectors 
defining the grid of deformed states of a reference 
state, and the corresponding Cauchy stress tensors 
calculated from DFT or derived by using a symmetry 
operation, the last operation consists in using 
Eq.\ \ref{nle} to transform Cauchy to 
PK2 stress tensor, and finally Eqs.\ \ref{opera} 
and \ref{cfd2a} to calculate all the components 
of the elastic tensors up to the selected order. 
\end{itemize}

\begin{figure}[!ht]
\centering
\includegraphics[width=\columnwidth]{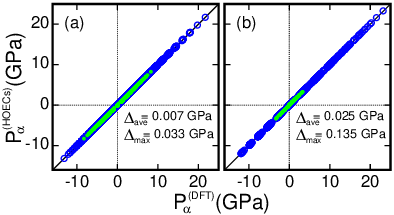}
\caption{\label{merit} 
Components of the PK2 stress tensor of the 
deformed configurations used to calculate the 
elastic constants up to the 6$^{th}$ of 
the (a) diamond phase of silicon and (b) {\it fcc} gold.
Values obtained by using Eq.\ \ref{pk2nu} 
are plotted against DFT results.
Normal and shear components are shown with 
blue and green circles, respectively. 
$\Delta_{ave}$ and $\Delta_{max}$ are 
the average and maximum absolute deviations 
between the two sets of values, respectively.}
\end{figure}

A Fortran module (``{\it hoecs.f90}'') performing the 
aforementioned operations is available under the GNU General 
Public License (Version 3) on GitHub \cite{mygit}.
This module can be compiled using the
standard GNU Fortran compiler, and it requires to 
be linked to BLAS and LAPACK libraries. Once compiled, 
its use mimics a standard Linux command accepting arguments. 
The first mandatory argument needs to be specified to 
select the operation to be performed: ``{\it str2pk}'', i.e. 
generation of the list of the deformed configurations of 
a reference state, or ``{\it pk2ecs}'', i.e. calculation 
of the elastic constants up to an order $n$. 
The module accepts additional optional arguments, to 
specify for example the working directory ({\it -d}), 
the filename containing information about the 
reference state ({\it -r}), the value of the 
strain parameter ({\it -s}), the highest order of the 
elastic constants to be calculated ({\it -n}),
and whether or not ({\it -nosym})
to exploit symmetry operations, or to impose or 
not ({\it -noperm}) permutational Voigt symmetry on 
the elastic tensors. For example, after successful 
compilation of the Fortran module to create the 
executable file {\it hoecs.x}, the following 
command-line instruction 
\begin{center}
{\it ./hoecs.x str2pk -r reference.dat -s 0.010 -n 5}
\end{center}
can be used to generate the list of deformed 
configurations of a reference state ({\it reference.dat}) 
necessary to calculate elastic constants up to 
the 5$^{th}$ order. Specifically, this command generates 
a list of files, two for each Lagrangian strain 
vector $\vec{\mu} = \xi \left( i , j , k, p , q, r \right)$
satisfying the condition $\left| i+j+k+p+q+r \right| \le 4$,
with strain parameter $\xi=0.010$. These files have 
default general names 
{\it ipos} and {\it istr}, followed by a six integer 
code extension, such as the pair {\it ipos.000001} 
and {\it istr.000001} referring to the reference 
state, or {\it ipos.002099} and {\it istr.002099} 
referring to an arbitrary deformed state. Unless the 
argument {\it -nosym} is present, the above 
command-line instruction generates a pair of {\it ipos} 
and {\it istr} files for each representative member 
of the families of deformed configurations 
equivalent by symmetry. The symmetry operations leaving 
invariant the reference state are identified by 
scanning the 32 possible symmetry elements of a crystal.
We remark that the current version of {\it hoecs.f90} 
is compatible with only the Quantum Espresso software 
package \cite{qea,qeb}. Thus, both  
{\it reference.dat} and the {\it ipos} files 
have the same format, compliant with the requirements 
of the Quantum Espresso software as to how to specify 
geometry of the unit cell and coordinates of the atoms. 
As for the {\it istr} files, each one contains
information about the Lagrangian strain vector 
$\vec{\mu}$ (i.e. $\xi$ and six integer coordinates) 
used to generate the deformed state, whose 
geometry is specified in the {\it ipos} file having 
the same six integer code extension. Furthermore, 
the {\it istr} files contain information about
the symmetry family each deformed state belongs to.

After generating the {\it ipos} and {\it istr} 
files via the command-line instruction above, 
a DFT calculation is carried out for 
each {\it ipos} file to optimize ionic position, 
release the internal stress arising from the 
deformation, and calculate the resulting 
stress tensor, which needs to be extracted from 
the output file and appended to the corresponding {\it istr} file. 
Execution of DFT calculations 
and subsequent file parsing operations to 
extract the components of the stress tensor 
are tasks that can be easily managed through 
the use of shell scripts and Unix commands.
After all DFT calculations are completed, 
the following command-line instruction
\begin{center}
{\it ./hoecs.x pk2ecs -d ./dft -matsym}
\end{center}
can be used to process all the
{\it ipos} and {\it istr} files contained in 
the directory {\it ./dft} (by default 
these files are expected to be located in 
the current working directory), and hence 
to calculate the elastic constants up to the 5$^{th}$ order. 
The option {\it -matsym} 
instructs the module to carry out the 
operations in Eq.\ \ref{sym}, and therefore 
to impose that the elastic tensors up to the 5$^{th}$ 
order are invariant under the symmetry 
operations of the reference state.
The Fortran module is complemented by several 
example applications, also available on GitHub \cite{mygit}. 
Each example includes the input files 
and shell scripts necessary to reproduce the 
results provided as reference, as well as 
README files detailing the sequence of 
instructions/tasks to complete the application, 
and providing information about the format of 
the {\it ipos} and {\it istr} files, and the 
output files containing the values of 
the elastic constants.

\begin{figure}[!ht]
\centering
\includegraphics[width=\columnwidth]{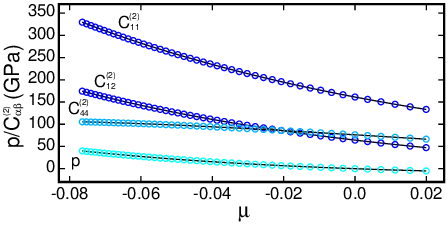}
\caption{\label{si.hyd} 
Pressure and independent SOECs 
of silicon versus hydrostatic (Lagrangian) strain.
DFT results (colored circles) are compared to 
values (black solid lines) obtained by using a 
nonlinear elastic constitutive framework 
(Eqs.\ \ref{nle} and \ref{pk2nu}) and 
higher-order elastic constants computed with the 
present method.}
\end{figure}

\subsection{Technical details of the DFT calculations}\label{dft}

We use the {\it pw.x} code of the Quantum ESPRESSO
package to carry out DFT calculations \cite{qea,qeb}.
All DFT calculations are carried out by employing
primitive unit cells, ultrasoft pseudopotentials, 
and strict convergence criteria (10$^{-14}$ Ry for 
self-consistency and 10$^{-6}$ {\it a.u.} for forces).
HOECs of silicon (space group $Fd\bar{3}m$) reported in 
the following tables 
are obtained by using a local density
approximation (LDA) functional \cite{pz81}, the pseudopotential
{\it Si.pz-nl-rrkjus\_psl.1.0.0.UPF} \cite{c14},
a uniform mesh of 10$\times$10$\times$10 $\bm{k}$-points to
sample the Brillouin zone, and
plane-wave energy cutoffs of 80 and 320 Ry
to represent wave functions and electron charge
density, respectively.
HOECs of gold (space group $Fm\bar{3}m$) are obtained 
by using a LDA functional \cite{pz81}, 
the pseudopotential {\it Au.pz-spn-rrkjus\_psl.1.0.0.UPF} \cite{c14}, 
a uniform mesh of 25$\times$25$\times$25 
$\bm{k}$-points, and energy cutoffs of 120 and 480 Ry.
To calculate HOECs of $\alpha$-quartz (space group P$3_121$ or P$3_221$),
we use a generalized gradient approximation
functional \cite{prc08}, the pseudopotentials 
{\it Si.pbesol-nl-rrkjus\_psl.1.0.0.UPF} and
{\it O.pbesol-nl-rrkjus\_psl.1.0.0.UPF} \cite{c14}, 
a uniform mesh of 6$\times$6$\times$6 $\bm{k}$-points,
and energy cutoffs of 100 and 400 Ry.
SOECs and TOECs of kevlar (space group P1n1) are 
obtained by using the nonlocal van der Waals density 
functional rVV10 \cite{vv90,sgg13}, 
a uniform mesh of 3$\times$3$\times$2 $\bm{k}$-points, 
energy cutoffs of 100 and 400 Ry, and the 
following ultrasoft pseudopotentials: 
{\it H.pbe-n-rrkjus\_psl.1.0.0.UPF},
{\it C.pbe-n-rrkjus\_psl.1.0.0.UPF},
{\it N.pbe-n-rrkjus\_psl.1.0.0.UPF}, and
{\it O.pbe-n-rrkjus\_psl.1.0.0.UPF} \cite{c14}.

The Supplementary Material reports on a thorough study 
to assess the convergence behavior of HOECs computed via 
our method. In this convergence study, we consider gold, 
a metal, and silicon, a semiconductor,
and we calculate elastic constants up to the 6$^{th}$ order 
using energy cutoffs and $\bm{k}$-point meshes of 
increasing values and density, respectively.
In brief, this study demonstrates that minimal 
(suggested in the pseudopotentials) energy cutoffs 
(of 56 and 399 Ry for gold, 
and 44 and 176 Ry for silicon) and conventional $\bm{k}$-meshes 
(of 25$\times$25$\times$25 for gold, and 
10$\times$10$\times$10 for silicon) yield converged 
values of SOECs, TOECs, and FOECs for both materials.
Energy cutoff larger than the minimal ones, 
and in case of a metal also denser $\bm{k}$-meshes, 
are needed to obtain a set of satisfactorily converged 
fifth- and sixth-order elastic constants, with 
small-valued HOECs converging, in general, more 
irregularly and carrying larger deviations than 
the large-valued ones. We remark that the small-value 
HOECs showing poor convergence are coefficients 
(such as $C^{(5)}_{11456}$ and $C^{(6)}_{112456}$ 
for gold, or $C^{(6)}_{114455}$ for silicon) 
controlling the stress response to deformations 
involving both normal and shear strains.

The following tables report elastic constants of 
gold, silicon, $\alpha$-quartz, and kevlar 
obtained by carrying out DFT calculations with 
energy cutoffs larger than the suggested ones and 
$\bm{k}$-meshes in the Brillouin zone of standard 
dimensions for a metal, a semiconductor, and 
insulators. Thus, based on the results reported in 
the Supplementary Material, elastic constants of these 
materials up to the 4$^{th}$ order are expected 
to be fully converged, whereas HOECs of the 5$^{th}$ 
and 6$^{th}$ order are expected to be satisfactorily 
well converged, with large-valued coefficients 
deviating on average by a few a percent points from 
fully converged values, and small-values constants 
carrying larger errors. Taking into consideration 
scope of this work and use made of the computed 
HOECs, the errors on (mainly the small-valued)  
5$^{th}$- and 6$^{th}$-order elastic constants are 
here considered as acceptable and tolerable.

\section{Results and discussion}\label{resdis}

\subsection{Silicon and gold}\label{siau}

We first use our method to calculate the elastic 
constants up to 6$^{th}$ order of Si and Au.
To accomplish this task, we first generate the grid of 
deformed configurations by using 
the following list of Voigt strain vectors: 
$\vec{\mu} = \xi \left( i , j , k, p , q, r \right)$,
with $ \left| i+j+k+p+q+r \right| \le 5$ and $\xi=0.015$.
For both solids, we consider a reference state 
at zero static pressure.
Then, we use a DFT approach \cite{qea,qeb} to 
calculate the PK2 stress tensor of 
the 297 (out of 3653) symmetry irreducible 
configurations, and finally 
Eqs. \ref{opera} and \ref{cfd2a} to 
calculate the elastic constants.
As figure of merit, Fig.\ \ref{merit} shows 
a comparison between components of the PK2 
stress tensor obtained from DFT calculations
and those derived from Eq.\ \ref{pk2nu}, 
and thereby from the computed higher-order 
elastic constants.
The excellent match between the two sets of values 
conveys two important messages. First, 
our method relying on recursive finite 
difference differentiation is sound. 
Second, the computed higher-order elastic constants 
encase information about the elastic behavior 
of these two materials over intervals of tens of GPa.

\begin{table*}[ht!]
\centering
\caption{\label{table1}
Independent SOECs and TOECs (in GPa) of Si and Au
obtained by using the present method.
Experimental data and previous computational results 
are also reported for comparison. 
For convenience, Voigt indices $\alpha\beta$ 
and $\alpha\beta\gamma$ are used to refer to 
$C^{(2)}_{\alpha\beta}$ and 
$C^{(3)}_{\alpha\beta\gamma}$, respectively. }
\begin{tabular}{c|ccc|cccccc}
\hline
 & 11 & 12 & 44 & 111 & 112 & 123 & 144 & 155 & 456 \\
 \hline
 \multicolumn{10}{c}{Silicon} \\
 \hline
 This work          & 161 & 64 & 76 & -769 & -459 & -87 & 31& -296 & -59 \\
 Exp.\ \cite{jh67}  & 166 & 64 & 80 & -795 & -445 & -75 & 15& -310 & -86 \\
 Ref.\ \cite{ccb18} & 142 & 51 & 72 & -744 & -393 & -59 &  4& -297 & -59 \\
 Ref.\ \cite{czl20} & 152 & 59 & 78 & -653 & -456 & -96 & 23& -304 & -7 \\
\hline
 \multicolumn{10}{c}{Gold} \\
\hline
 This work              & 214 & 187 & 36 & -2026 & -1219 & -369 & -43 & -781 & 74\\
 Exp.\ \cite{na58,hg66} & 192 & 163 & 42 & -1729 &  -922 & -233 & -13 & -648 & -12 \\
 Ref.\ \cite{pb23cpc}   & 207 & 179 & 35 & -1985 & -1177 & -373 & -63 & -749 & 63 \\
 Ref. \cite{wl09}       & 202 & 174 & 38 & -2023 & -1266 & -263 & -63 & -930 & 54 \\
 Ref.\ \cite{llz22}     & 151 & 126 & 28 & -1438 &  -875 & -550 & -66 & -469 & 16 \\
\hline
\end{tabular}
\end{table*}

Table \ref{table1} reports the independent 
SOECs and TOECs of Si and Au obtained in this work.
These results are in excellent agreement 
with both available experimental \cite{jh67,na58,hg66}
and computational data \cite{wl09,czl20,llz22,ccb18,pb23cpc}.
The small differences between our 
results and the experimental data are attributed to 
the following factors.
Our method yields elastic constants of a perfect 
monocrystalline material at zero temperature, whereas measurements 
are carried out at room temperature on polycrystalline samples 
containing defects, which are known to affect experimental 
data \cite{wl09,tek17}. 
Also, while our results are sensitive to the 
technical details of the DFT calculations, the analysis 
of the experimental data relies on approximations that 
have been shown to be a potential source of 
errors \cite{myy21}.
All in all, the results reported in Table \ref{table1} 
show that the present method is a valid alternative to 
available techniques to calculate TOECs.

Table \ref{table2} reports the 11 independent 
FOECs \cite{t24} of Si and Au obtained using our 
method, together with previous computational 
results \cite{wl09,czl20,llz22,pb23cpc}.
Our results for Au are in agreement with the values 
obtained by Pandit {\it et al.} \cite{pb23cpc} 
and Wang {\it et al.} \cite{wl09}, whereas 
differences exist with respect to 
data reported by Liao {\it et al.} \cite{llz22} (Table \ref{table2}).
However, it is to be noted that SOECs of Au reported 
in this last work \cite{llz22} 
deviate by 20-30\% from experimental values, thereby 
suggesting that their HOECs
may carry similar or larger errors.
As for Si, our FOECs agree with the results by 
Pandit {\it et al.} \cite{pb23cpc}, whereas 
they differ from the values reported in Ref.\ \cite{czl20}.
In this case, it is arguable that these 
differences arise from 
the strategy used to calculate the elastic constants,
markedly different from ours. 
In particular, while our method relies on a recursive 
numerical differentiation technique, in Ref.\ \cite{czl20} 
elastic constants were derived by carrying out 
a nonlinear least-square regression of a large dataset 
of stress and energy values, generated by 
considering deformations of a reference state 
spanning wide intervals of strain.

\begin{table*}[ht!]
\centering
\caption{\label{table2}
Independent FOECs (in GPa) of silicon and gold obtained 
by using the present method. Our results are compared 
to available experimental and computational data. 
Elastic constants are referred to by using quadruplets 
of Voigt indices.}
\begin{tabular}{c|ccccccccccc}
\hline
                   & 1111 & 1112 & 1122 & 1123 & 1144 & 1155 & 1255 & 1266 & 1456 & 4444 & 4455\\
\hline
 \multicolumn{11}{c}{Silicon} \\
\hline
This work           &2559 & 2246 & 2072 &  618 & -776 & 854 & -512& 732& -76 & 1343 & 15 \\
Ref.\ \cite{pb23cpc}&2586 & 2112 & 1885 &  576 & -671 & 833 & -422& 742& -46 & 1268 & -2  \\
Ref.\ \cite{czl20}  & 613 & 2401 & 1275 & 1053 & 5071 &4050 &-2728&-514&  66 &-2553 &-577 \\
\hline
 \multicolumn{11}{c}{Gold} \\
\hline
	This work   & 17718 & 8150& 9269 & 910 &1236 & 7854& -895& 7657&-228& 9204&-58\\
Ref.\ \cite{pb23cpc}& 17113 & 8114& 8814 & 874 & 860 & 7462& -634& 7372&-257& 8258&-61 \\
Ref.\ \cite{wl09}   & 17951 & 8729& 9033 & 416 & 691 & 7774& -752& 9402&-170& 8352& 15 \\
Ref.\ \cite{llz22}  & 10094 & 8280& 8402 &1507 & 235 & 5549&-1534& 8252&   2& 3640&-5763 \\
\hline
\end{tabular}
\end{table*}

\begin{figure}[!ht]
\centering
\includegraphics[width=\columnwidth]{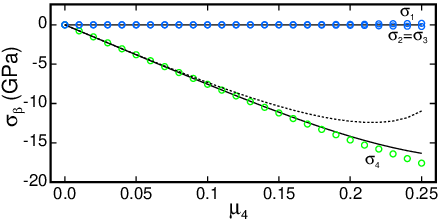}
\caption{\label{si.she} Components of the 
Cauchy stress tensor of Si versus shear strain in the $yz$ plane. 
Colored circles show DFT results, whereas black lines show 
values derived from nonlinear elastic 
constitutive equations (Eqs.\ \ref{nle} and \ref{pk2nu}) 
employing elastic constants up to the 6$^{th}$ (solid) 
and 3$^{rd}$ order (dashed). }
\end{figure}

\begin{figure}[!ht]
\centering
\includegraphics[width=\columnwidth]{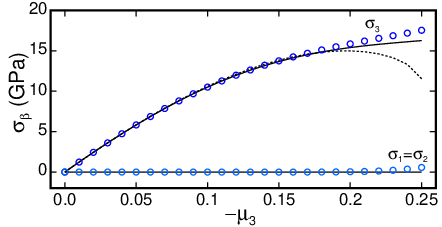}
\caption{\label{si.uni} Components of the 
Cauchy stress tensor of Si versus normal strain 
along the $z$ axis. Symbols and lines as 
in Fig.\ \ref{si.she}.}
\end{figure}

\begin{figure}[!ht]
\centering
\includegraphics[width=\columnwidth]{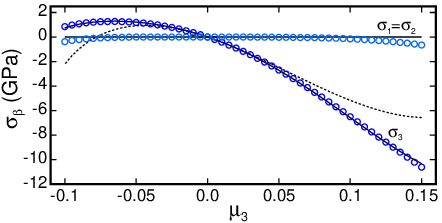}
\caption{\label{au.uni} Components of the
Cauchy stress tensor of Au versus normal strain
along the $z$ axis. Symbols and lines 
as in Fig.\ \ref{si.she}.}
\end{figure}

\begin{figure}[!ht]
\centering
\includegraphics[width=\columnwidth]{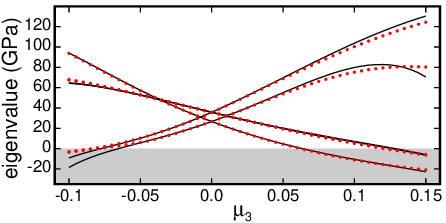}
\caption{\label{au.sta} Lowest 
five eigenvalues of the Birch tensor of 
{\it fcc} gold versus normal strain along 
the $z$ direction. Stress components $xx$ and $yy$ 
are zero. Red discs show DFT results, whereas 
solid black lines show results obtained from 
nonlinear elastic constitutive equations employing
elastic constants up to the 6$^{th}$ order. The 
shaded area highlights the region of mechanical 
instability.}
\end{figure}

The 18 and 32 independent fifth- and sixth-order elastic 
constants of Si and Au are reported in 
tables \ref{table3} and \ref{table4}, respectively. 
These HOECs have been also calculated using larger 
energy cutoffs and denser $\bm{k}$-meshes (see Supplementary 
Material).
To assess the fairness of the HOECs computed 
with the technical specification in Sec.\ \ref{dft}, 
we explore the nonlinear elastic responses
of Si and Au (treated as continuum elastic media)
over large intervals of strain. To this end, 
we employ nonlinear elastic constitutive 
equations (Eqs.\ \ref{nle} and \ref{pk2nu}) 
combined with optimization algorithms based 
on the Newton method \cite{bb22,pb23prm}.
We first consider Si under hydrostatic pressure. 
In detail, we use our numerical 
framework based on Eqs.\ \ref{nle} and \ref{pk2nu} 
to compress hydrostatically the reference state 
from 0 up to 40 GPa, at intervals of 1 GPa (Fig.\ \ref{si.hyd}). 
At each step, we determine the strain required 
to achieve the compressed state, and the 
corresponding second-order elastic tensor. 
The deformed geometries of the reference state 
generated by our nonlinear elastic constitutive 
framework are then provided to our DFT 
approach to calculate both stress tensor 
and SOECs. The two sets of results are compared 
in Fig.\ \ref{si.hyd}. 
Second, we consider Si subjected to, separately, 
uniaxial and shear deformations. In particular, 
we consider compression along the $z$ axis 
and shearing in the $yz$ plane, spanning the 
strain intervals shown in 
Figs.\ \ref{si.she} and \ref{si.uni}, respectively.
At each step of the deformations, axial ($\mu_3$) 
and shear ($\mu_4$) strains are reduced and 
incremented by 0.01, respectively. The numerical 
framework relying on HOECs is then 
used to determine the geometry of the 
deformed states yielding zero stress 
components, except for $\sigma_3$ and $\sigma_4$, 
respectively. Also in this case, these deformed 
geometries are supplied to our DFT approach to 
calculate the stress tensor. The set of relevant 
stress tensor components obtained from these 
calculations are compared in 
Figs.\ \ref{si.she} and \ref{si.uni}.
Third, we consider Au subjected to uniaxial strain. 
In this case, $\mu_3$ is varied from -0.10 to 0.15 
at intervals of 0.01. At each step, first we determine 
the geometry of the deformed state yielding zero stress  
components (except $\sigma_3$) while 
keeping the value of $\mu_3$ fixed. Then, 
we calculate the symmetrized second-order Birch 
tensor \cite{w67,bk65} and the corresponding eigenvalues and 
eigenvectors, carrying information 
about the mechanical stability of the deformed Au lattice. 
Also in this case, the 
deformed geometries produced by 
the nonlinear elastic constitutive framework are used to 
calculate stress and Birch tensors directly 
by using the DFT approach. These results are compared in 
Figs.\ \ref{au.uni} and \ref{au.sta}.

\begin{table}[ht!]
\centering
\caption{\label{table3}
Independent fifth-order elastic 
constants (in GPa) of Si and Au obtained 
by using the present method.
Voigt indices are used to refer to respective 
elastic constants.}
\begin{tabular}{c|r|r||c|r|r}
\hline
$C^{(5)}_{ijklm}$ & Si & Au & $C^{(5)}_{ijklm}$ & Si & Au \\
\hline
11111 & -7258& -191822 & 11266 & -2900 & -66307\\
11112 &-10297&  -58296 & 11456 &  2078 &    809\\
11122 &-10782&  -72938 & 12344 &  4909 &   6807\\
11123 & -3578&    8649 & 12456 &  -827 &   1067\\
11144 &  8473&  -17505 & 14444 &  1043 &  -9110\\
11155 & -4991&  -74910 & 14455 &  -352 &    -30\\
11223 & -2858&   -6075 & 15555 & -7423 & -69308\\
11244 &  3128&    2269 & 15566 &  -883 &   -620\\
11255 &  2701&    7686 & 44456 &  2416 &   5377\\
\hline
\end{tabular}
\end{table}

\begin{table*}[ht!]
\centering
\caption{\label{table4}
Independent sixth-order elastic 
constants (in GPa) of Si and Au obtained 
by using the present method.}
\begin{tabular}{l|r|r||l|r|r}
\hline
	$C^{(6)}_{ijklmn}$ & Si & Au & $C^{(6)}_{ijklmn}$ & Si & Au \\
\hline
 111111& 125882&  2824259& 112355& -24696&   139770\\
 111112&  24814&   844342& 112456&   7508&    -2349\\
 111122&  52871&   495330& 114444&  52563&  -713028\\
 111123&  45832&  -259263& 114455&    936&    67971\\
 111144& -67645&  -107926& 115555&  85601&   118851\\
 111155&  55610&   716705& 115566&   2251&   146435\\
 111222&  90551&  1021102& 123456&   2948&    17858\\
 111223&   7549&    31322& 124444&  17270&   692953\\
 111244& -14844&   314343& 124455&   9013&      438\\
 111255&   8625&   206475& 124466&   2400&    90682\\
 111266&  50950&   651330& 126666&  65230&   101842\\
 111456&   7186&    77641& 144456&  13595&     4738\\
 112233&  31225&    80440& 145556&  12239&    27293\\
 112244&  -7366&  -254564& 444444&  62603& -1349191\\
 112266&  36570&   450821& 444455& -37457&   129888\\
 112344& -20967&   -87887& 445566& -52477&     2895\\
\hline
\end{tabular}
\end{table*}

The comparisons in Figs.\ \ref{si.hyd}-\ref{au.sta} 
show that HOECs computed via the present method can 
be used to reproduce the elastic behavior of 
a material over finite intervals of strain and
stress around a reference state.
The intervals of strain and 
stress over which the nonlinear elastic model 
reproduces DFT results broaden with increasing 
the order of the polynomial in Eq.\ \ref{nle}, 
from about $\pm$0.02 and $\pm$5 GPa when 
only SOECs and TOECs 
are accounted for, up to $\pm$0.2 and $\pm$20 GPa when 
elastic constants up to the 6$^{th}$ order 
are used (Figs.\ \ref{si.she}-\ref{au.uni}). 
These results constitute an 
indirect demonstration that the present method 
yields reliable values of HOECs. 
Furthermore, they show that HOECs encode 
information about the elastic behavior of a material, 
and they can be used to pinpoint and characterize the 
elastic limits and failure mechanisms of a lattice. 
For example, in agreement with previous
computational studies \cite{wl10}, our 
results in Fig.\ \ref{au.sta} 
show that HOECs (up to the 6$^{th}$ order) 
correctly predict that upon uniaxial tension, 
{\it fcc} gold fails at around a strain of 0.06 and 
a tensile stress of 3.5 GPa, when the stability 
condition, $C^{(2)}_{11}-C^{(2)}_{12}>0$, 
ceases to be valid, triggering 
symmetry breaking (bifurcation) transformations 
by tetragonal shearing deformations. 

\begin{table}[ht!]
\centering
\caption{\label{table5}
Independent SOECs 
of $\alpha$-SiO$_2$ at 0, 10, and 35 GPa obtained by using 
the present method. Voigt indices are 
used to refer to elastic coefficients.} 
\begin{tabular}{c|cccccc}
 \hline
 $p$ & 11 & 12 & 13 & 14 & 33 & 44 \\
 \hline
 0 &  75&  1 & 6  & 19 & 88  & 52 \\
10 & 120& 43 & 43 & -2 & 209 & 66 \\
35 & 269& 85 & 77 &-50 & 404 & 87 \\
 \hline
\end{tabular}
\end{table}

\begin{table*}[ht!]
\centering
\caption{\label{table6}
Independent TOECs
of $\alpha$-SiO$_2$ at 0, 10, and 35 GPa obtained by using
the present method. Voigt indices are 
used to refer to elastic coefficients.}
\begin{tabular}{c|cccccccccccccc}
 \hline
 $p$ & 111 & 112 & 113 & 114 & 123 & 124 & 133 & 134 & 144 & 155 & 222 & 333 & 344 & 444 \\
 \hline
 0 &     3 & -339&  86 & 123 &-277 & 24 & -224& -26 & -186 & -164 & -98 & -623 &-124 & 220\\
10 &  -694 & -323& -47 & 372 &-248 &-57 & -489&  55 & -117 &  -57 &-913 &-2267 &-79  & 110\\
35 & -2348 & -799&-426 & 823 & -29 & 36  & 2  &   69& -62 & -298 & -3035&-3752&  62  & 176 \\
 \hline
\end{tabular}
\end{table*}

\subsection{$\alpha$-quartz}\label{quar}

To demonstrate potential 
applications of our method, we first focus 
on $\alpha$-quartz, a crystalline material 
with the trigonal symmetry undergoing a 
pressure-induced amorphization transition at 
$>$20 GPa \cite{kmh93,khm93,ghm00,wmf15,hsy17,bc92,cc06,kol07,bqj24}.
The crystalline phase of $\alpha$-quartz 
loses mechanical stability when at least one the following 
conditions fail to be satisfied \cite{ghm00,ms03,ghm03}:
\begin{equation}\label{stcon}
\begin{split}
&K0 = B_{44} > 0  \\
&K1 = B_{11} - \left| B_{12}\right| > 0 \\
&K2 = \left( B_{11} + B_{12} \right) B_{33} - 2 [B_{13}]^2 > 0 \\
&K3 = B_{44} B_{66} - [B_{14}]^2 > 0,
\end{split}
\end{equation}
where $B_{\alpha\beta}$ are Birch coefficients \cite{w67,bk65}, 
which at a hydrostatic pressure $p$ are defined as, 
\begin{equation}\label{birch}
B_{ijkl}= C^{(2)}_{ijkl} + p \left ( \delta_{ij} \delta_{kl}  -
        \delta_{ik} \delta_{jl}  - \delta_{il} \delta_{jk} \right ),
\end{equation}
where $C^{(2)}_{ijkl}$ are 
SOECs defined as in Eq.\ \ref{noecs} and tensor 
components are expressed in terms of cartesian indices.
Here, we use our method to calculate the elastic constants 
of $\alpha$-quartz up to the 5$^{th}$ order, and then 
we use a nonlinear elastic constitutive framework based 
on Eqs.\ \ref{nle} and \ref{pk2nu} to chart the linear 
elastic properties and assess 
mechanical stability over an interval of pressure. 
In particular, to obtain a comprehensive map of 
the elastic behavior of $\alpha$-quartz, from 
zero pressure up to its stability limit, we consider 
three reference states, yielding a static pressure 
of 0, 10, and 35 GPa, respectively.
For each reference state, we use 
(Voigt) strain vectors  $\vec{\mu} = \xi \left( i , j , k, p , q, r \right)$
with $ \left| i+j+k+p+q+r \right| \le 4$ and $\xi=0.010$ to 
generate the 1289 deformed configurations, 
and then we use a DFT approach \cite{qea,qeb} to
calculate the stress tensor of the 805 members 
that are irreducible by symmetry. These DFT calculations 
are carried out using the same technical details as in 
Ref. \cite{pb23prm}, which have been 
shown to yield a satisfactory description of 
SOECs and TOECs of $\alpha$-quartz over a wide 
range of pressures \cite{pb23prm}. Tables \ref{table5} 
and \ref{table6} report the 6 and 14 independent SOECs 
and TOECs of $\alpha$-SiO$_2$ at 0, 10, and 35 GPa, 
respectively. The 28 and 52 indepependent fourth- and  
fifth-order elastic contants 
are reported in tables \ref{table7} and 
\ref{table8}, respectively.


\begin{table}[ht!]
\centering
\caption{\label{table7}
Independent fourth-order elastic
constants (in GPa) of $\alpha$-SiO$_2$ at 0, 10, and 35 GPa 
obtained by using the present method. }
\begin{tabular}{c|rrr}
\hline
$C^{(4)}_{ijkl}$ & 0 & 10 & 35 \\
\hline
 1111 &    15107 &     8881 &    17480 \\
 1112 &      361 &       56 &    14746 \\
 1113 &     2458 &     -568 &    12101 \\
 1114 &    -7621 &    -3379 &    -7687 \\
 1122 &    -2172 &      149 &     4615 \\
 1123 &      141 &      694 &      561 \\
 1124 &     1358 &     -135 &    -2606 \\
 1133 &     1625 &    -3947 &    -2090 \\
 1134 &    -4299 &     -104 &    -2313 \\
 1144 &      765 &       77 &    -1349 \\
 1155 &    -3427 &     1822 &     6078 \\
 1156 &     -176 &    -1310 &    -5515 \\
 1166 &     2825 &     3149 &     9351 \\
 1223 &    -2882 &    -1647 &      536 \\
 1233 &      926 &     1961 &    -2060 \\
 1234 &      699 &      456 &      433 \\
 1244 &    -3655 &      -49 &      638 \\
 1255 &    -2483 &     -537 &     -218 \\
 1333 &     2963 &     6204 &    -6940 \\
 1334 &    -1695 &      472 &     1373 \\
 1344 &    -1034 &     1869 &    -2199 \\
 1355 &    -2737 &      684 &    -3499 \\
 1444 &     1471 &      593 &      165 \\
 1455 &     -891 &      266 &      725 \\
 3333 &     9663 &    30385 &    40309 \\
 3344 &    -1559 &     1013 &    -3697 \\
 3444 &      171 &     -176 &    -2978 \\
 4444 &    -1573 &     -351 &    -3120 \\
\hline
\end{tabular}
\end{table}

\begin{figure}[!ht]
\centering
\includegraphics[width=\columnwidth]{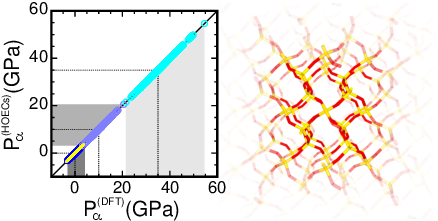}
\caption{\label{qua.mer} 
Components of the PK2 stress 
tensor used to calculate the elastic constants 
up to 5$^{th}$ order of $\alpha$-quartz 
in three different reference states, yielding 
static pressures of 0, 10, and 35 GPa, respectively.
Values obtained by using Eq.\ \ref{pk2nu}
are plotted against DFT results.
Shear components are shown in yellow, 
whereas blue, purple, and cyan 
colored circles show normal components of the 
PK2 stress tensor arising in deformed configurations of 
the reference states at 0, 10, and 35 GPa, respectively.
Shaded areas show the intervals of normal 
stresses spanned by Eq.\ \ref{pk2nu} 
around each reference state. Left panel, 
image showing the 
bond network of $\alpha$-quartz.}
\end{figure}

\begin{figure}[!ht]
\centering
\includegraphics[width=\columnwidth]{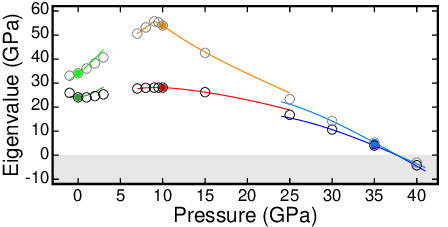}
\caption{\label{qua.sta} 
Lowest two eigenvalues of the Birch tensor of 
$\alpha$-quartz versus pressure. Circles show 
results obtained from DFT, whereas solid lines 
show values derived from nonlinear elastic 
constitutive equations employing
elastic constants up to 5$^{th}$ order for 
the reference state at 0 (dark and light green), 
10 (red and orange), and 35 GPa (blue 
and light blue), respectively. 
The shaded area shows the region of mechanical
instability.}
\end{figure}

\begin{figure}[!ht]
\centering
\includegraphics[width=\columnwidth]{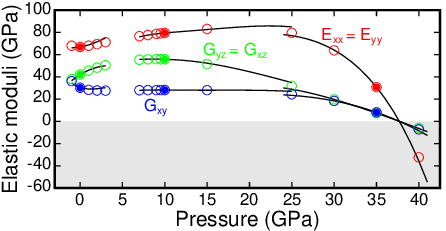}
\caption{\label{qua.eg} 
Selected elastic moduli of $\alpha$-quartz versus 
pressure. Circles show DFT results, 
whereas black solid lines show values derived 
from constitutive equations using 
elastic constants up to 5$^{th}$ order of the 
reference state at 0 (left),
10 (middle), and 35 GPa (right), respectively.
Red circles show values of the 
Young's modulus along a direction perpendicular to
the 3-fold symmetry $c$-axis of $\alpha$-quartz, 
whereas green and blue circles show 
the shear moduli associated with planes parallel 
and perpendicular to the $c$-axis, respectively. }
\end{figure}

In Fig.\ref{qua.mer}, PK2 stress components of 
the 805 deformed states 
obtained from DFT calculations are compared to 
results derived from Eq.\ \ref{pk2nu} 
by using HOECs calculated using our 
method (Eqs. \ref{opera} and \ref{cfd2a}).
As discussed above (Fig.\ \ref{merit}),
this comparison demonstrates 
that our method to calculate 
HOECs is sound, and that the computed nonlinear 
elastic constants encode information about the 
elastic behavior of $\alpha$-quartz over finite 
intervals of pressure.
In detail, Fig.\ref{qua.mer} shows that
elastic constants up to the 5$^{th}$ order 
can be used to predict trends in elastic 
properties over an interval 
of several GPa at zero pressure, and 
up to about 15 and 40 GPa as $\alpha$-quartz 
stiffens at around the reference pressures of 
10 and 35 GPa, respectively. This is clearly shown in 
Figs. \ref{qua.sta} and \ref{qua.eg}, 
reporting results obtained by using a
nonlinear elastic constitutive framework 
(Eqs.\ \ref{nle} and \ref{pk2nu}) based on 
HOECs computed at the three reference pressures 
of 0, 10, and 35 GPa. In particular, 
Fig.\ \ref{qua.sta} shows the lowest two 
eigenvalues of the Birch tensor (Eq.\ \ref{birch}) 
from 0 up to about 40 GPa, whereas 
Fig.\ \ref{qua.eg} shows the shear moduli and 
the Young's modulus along a direction 
perpendicular to the 3-fold symmetry c-axis 
$\alpha$-quartz. For completeness, 
and to further validate our method to calculate 
HOECs, these predictions are confronted to 
results obtained from explicit 
DFT calculations. Overall, in agreement with 
previous computational studies \cite{bc92}, our results 
show that the amorphization transition occurs 
at a pressure of about 37 GPa, and that loss of 
mechanical stability at this pressure arises 
from failure of condition $K3$ in Eq.\ \ref{stcon}.
Contrary to claims of early experimental 
studies \cite{ghm00,ms03}, although in agreement with more recent 
experiments and computational studies \cite{wmf15,bc92,bqj24}, our 
results show that condition $K0$ in Eq.\ \ref{stcon}, 
i.e. the value of $B_{44}=C^{(2)}_{44} - p$, 
remains positive at and beyond the transition 
pressure, along with the other two stability 
conditions $K1$ and $K2$. Interestingly, Fig.\ \ref{qua.eg} 
shows that mechanical instability at about 37 GPa 
coincides with the combined failure of all 
shear moduli, along with the Young's moduli along 
directions perpendicular to the $c$-axis of 
$\alpha$-quartz, which might explain why the 
mechanical instability leads to the amorphization 
of the crystalline phase, rather than a 
transition to a high pressure crystalline phase 
of SiO$_2$.

\begin{table}[ht!]
\centering
\caption{\label{table8}
Independent fifth-order elastic
constants (in GPa) of $\alpha$-SiO$_2$ at 0, 10, and 35 GPa
obtained by using the present method. }
\begin{tabular}{c|rrr}
\hline
$C^{(5)}_{ijklm}$ & 0 & 10 & 35 \\
\hline
 11111 &   897619 &  -520028 &    84189 \\
 11112 &   146638 &    35285 &  -217615 \\
 11113 &   313965 &   -69950 &  -201915 \\
 11114 &  -900127 &  -134663 &    40667 \\
 11122 &    29670 &   -55958 &    64182 \\
 11123 &   -88019 &   -20458 &    35240 \\
 11124 &   242403 &   160768 &   157895 \\
 11133 &   181971 &  -218869 &  -257921 \\
 11134 &   -44731 &   -13779 &   -17764 \\
 11144 &   340860 &    -2195 &    20238 \\
 11155 &    42431 &  -217155 &  -346452 \\
 11156 &  -246322 &    -7576 &   138672 \\
 11166 &   201649 &   -83811 &  -156234 \\
 11222 &  -147834 &    34819 &  -245552 \\
 11223 &   -69967 &   -40289 &  -111479 \\
 11224 &  -131069 &  -108619 &  -208611 \\
 11233 &     5523 &    -8593 &     8266 \\
 11234 &    83771 &    37592 &   -24253 \\
 11244 &  -116345 &    39712 &  -101926 \\
 11255 &    19554 &   153758 &    51674 \\
 11256 &    -6501 &    47496 &   -45379 \\
 11333 &    39084 &    66328 &    60860 \\
 11334 &  -222845 &   -35662 &   -19343 \\
 11344 &   107581 &    56865 &   -45423 \\
 11355 &   -26112 &     1004 &   -53533 \\
 11356 &   -38832 &    -8935 &    24370 \\
 11366 &    64722 &   -22392 &   -17382 \\
 11444 &  -348880 &   101788 &  -112963 \\
 11455 &  -142807 &    -6502 &    38522 \\
 11456 &   -13498 &   -41709 &   -83139 \\
 12233 &   -29185 &    13682 &   -27752 \\
 12244 &   -44729 &   -22624 &   147168 \\
 12333 &  -111653 &   -61677 &    78164 \\
 12334 &    -6921 &    -1241 &    19008 \\
 12344 &  -136988 &      -43 &    45386 \\
 12355 &  -123100 &     7457 &    22046 \\
 12444 &    79095 &   -82473 &   164507 \\
 12455 &   -22366 &    16139 &    -6602 \\
 13333 &   -70335 &  -172528 &    53677 \\
 13334 &   -68899 &   -43476 &   -79335 \\
 13344 &    25346 &   -31264 &   -20991 \\
 13355 &   -46919 &   -39145 &     5674 \\
 13444 &  -110518 &     2968 &    -4171 \\
 13455 &   -53637 &   -26420 &   -57788 \\
 14444 &    64469 &   -18683 &    96856 \\
 14455 &    12995 &    26543 &   -85279 \\
 15555 &   -72374 &   -41779 &  -170959 \\
 33333 &  -171725 &  -360236 &  -567293 \\
 33344 &    16326 &    31309 &    27627 \\
 33444 &      586 &    20246 &    84045 \\
 34444 &    36463 &    72777 &    -5327 \\
 44444 &   370158 &    35859 &  -128381 \\
\hline
\end{tabular}
\end{table}

\subsection{Kevlar}\label{kevlar}

To demonstrate broad applicability of our method, 
we consider the case of kevlar, a man-made polymer 
fiber with an anisotropic bonding network exhibiting 
remarkable mechanical properties. 
This material is made of long chains of 
poly($p$-phenylene terephthalamide) (PPTA) 
monomers, aligned along one direction, interacting via 
hydrogen bonds and van der Waals interactions in the 
transverse direction, and arranged in crystalline domains 
exhibiting the orthogonal symmetry \cite{n74}. 
Although microstructure, defects, and 
fiber morphology play a major role \cite{yld92}, understanding 
the linear and nonlinear elastic properties of the 
crystalline domains is the first step towards elucidating 
and controlling the mechanical properties of this complex 
material. To this end, here we use our method to
calculate SOECs and TOECs of PPTA.
To the best of our knowledge, while linear elastic moduli 
of kevlar are known and have been calculated by using both 
force-field \cite{mzp17} and DFT methods \cite{yfp20}, no 
experimental and computational data are available for 
the nonlinear elastic properties of this material. 

PPTA has a crystalline structure with a monoclinic
(pseudo-orthorhombic) unit cell \cite{n74} containing 
two chains with chemical formula 
[-CO-C$_6$H$_4$-CO-NH-C$_6$H$_4$-NH-]$_n$, aligned along 
the $c$ axis. A DFT geometry optimization calculation of 
the unit cell of PPTA (space group number 7, P1n1) \cite{n74} 
gives the following lattice parameters: $a$=7.45 \AA,
$b$=5.15 \AA, $c$=13.10 \AA, and $\gamma$=90.014$^{\circ}$.
These values compare well with both 
experimental data \cite{n74} 
($a$=7.87 \AA, $b$=5.18 \AA, $c$=12.9 \AA, 
and $\gamma\simeq90^{\circ}$) and results of a recent 
DFT study \cite{yfp20}. It is interesting to note that 
the parameter deviating the most from the experiment
is $a$, the direction governed by 
van der Waals interactions. A recent DFT study 
employed several exchange and correlation energy 
functionals to describe PPTA, reporting values for 
$a$ ranging from 7.21 to 9.71 \AA \cite{yfp20}. 
This study highlighted the importance of using an 
exchange and correlation energy functional accounting 
for van der Waals interactions to describe PPTA. 
Here, we use the rVV10 functional \cite{vv90,sgg13}, 
designed specifically to reproduce 
equilibrium intermonomer separations in van der Waals 
complexes.

\begin{table*}[ht!]
\centering
\caption{\label{table9}
Independent SOECs and TOECs of kevlar at 0 GPa obtained by using
the present method. }
\begin{tabular}{ccccc|ccccccccc}
\hline
11 & 12 & 13 & 33 & 44 & 
	111 & 112 & 113 & 123 & 133 & 144 & 155 & 333 & 344 \\
\hline
34 & 16 & 4 & 339 & 10 & 
       -542 & -142 & -160 & -45 & -240 & -45 & -74 & -2100 & -57 \\
 \hline
 \hline
\end{tabular}
\end{table*}




We use our method to calculate SOECs and TOECs of kevlar, 
as described using the aforementioned monoclinic
(pseudo-orthorhombic) unit cell. In detail, we use a 
strain parameter of 0.010 to generate the deformed 
configurations and obtain the elastic constants of the monoclinic 
system. To obtain SOECs and TOECs of kevlar treated as 
a transverse isotropic material, we then carry out 
the following averaging operation \cite{rb24}:
\begin{equation}\label{isosym}
        \begin{split}
\overline{C}^{(2)}_{ijkl} &= \frac{1}{N} \sum_{\theta=1}^N
a^{(\theta)}_{ip} a^{(\theta)}_{jq} a^{(\theta)}_{kr}
a^{(\theta)}_{ls} C^{(2)}_{pqrs} \\
\overline{C}^{(3)}_{ijklmn} &= \frac{1}{N} \sum_{\theta=1}^N
a^{(\theta)}_{ip} a^{(\theta)}_{jq} a^{(\theta)}_{kr}
a^{(\theta)}_{ls} a^{(\theta)}_{mt} a^{(\theta)}_{nu}
C^{(3)}_{pqrstu},
        \end{split}
        \end{equation}
where $C^{(2)}_{pqrs}$ and $C^{(3)}_{pqrstu}$ are
SOECs and TOECs of the monoclinic material,
$\overline{C}^{(2)}_{ijkl}$ and
$\overline{C}^{(3)}_{ijklmn}$ are the symmetrized
elastic constants, and $a^{(\theta)}_{ij}$ is 
a 3$\times$3 matrix performing a rotation around the $c$ 
axis by a degree angle $\theta$, ranging from 1 and 
$N=360$. 

The independent SOECs and TOECs of kevlar 
obtained by carrying out the operations in Eq.\ \ref{isosym} 
are reported in Table \ref{table9}.
Overall, our SOECs agree well with 
results obtained from molecular dynamics simulations at 300 K 
based on force fields \cite{mzp17}. In particular, we find 
a Young's modulus parallel to the chain direction of 339 GPa,
in good agreement with values obtained by using force-field 
methods, ranging from 291 and 336 GPa \cite{mzp17}, 
and DFT calculations, ranging from 296 and 316 \cite{yfp20}.
As for TOECs, the values in Table \ref{table9} 
show the expected trends \cite{pb23prm}, with 
$\overline{C}^{(3)}_{333}$ and $\overline{C}^{(3)}_{111}$ 
exhibiting the largest negative values.

\section{Conclusions}\label{conc}

We introduced a first-principles method to calculate 
higher-order elastic constants of a solid material.
Our method relies on the use of a uniform grid of 
finite strain deformations of a reference state, 
a DFT approach to calculate the PK2 stress tensor, 
material symmetries to reduce the number of 
DFT calculations, and a recursive numerical 
differentiation technique homologous to the divided 
differences polynomial interpolation algorithm to compute the 
elastic tensors up to a order $n \ge 2$.
The method is general and applicable as is to any 
material, irrespective of its symmetry, to calculate 
elastic constants of, in principle, any order.
This method suffers from two types of errors. 
First, truncation errors, intrinsic and stemming from 
to the use of numerical differentiation techniques. 
These errors are minimal and can be reduced to 
less than a few percent of the 
values of the elastic constants by 
using strain parameters smaller than 0.015.
Second, errors extrinsic to the nature of the 
method, stemming from the nonuniform numerical 
precision of the stress tensor computed 
via DFT across the grid of deformed states.
These errors, which can be mitigated by increasing 
the numerical accuracy of the DFT calculations, 
are typically small for elastic constants 
up to the 4$^{th}$ order, and they gradually 
increase for elastic constants of higher order.
High-precision DFT calculations need to be 
carried out to obtain full convergence on 
the whose set of fifth-, sixth-, or higher-order 
elastic constants.
In this work, our method is applied to 
silicon, gold, $\alpha$-quartz, and kevlar. 
Besides demonstrating validity and 
novelty, these applications show that 
our method to calculate HOECs promises to 
enable a plethora of fundamental studies of 
materials under hydrostatic and/or non-hydrostatic 
stress conditions, and/or undergoing structural phase 
transitions.

\section{Declaration of competing interest}

The authors declare that they have no known competing financial 
interests or personal relationships that could have appeared to
influence the work reported in this paper.

\section{Supplementary material}

The supplementary material related to this article reports on 
a convergence study assessing the dependence of HOECs computed 
via our method on the accuracy of plane-wave-based DFT calculations.
This convergence study can be found online at ...

\section{Data availability}

Codes resulting from this work are available on GitHub
(https://github.com/abongiox/hoecs).

\section{Acknowledgements}

This work is supported by the National Science
Foundation (NSF), Award No. DMR-2036176.
We also acknowledge the support of the CUNY
High Performance Computing Center
(Award No. OAC-2215760).

\clearpage
\bibliographystyle{elsarticle-num}

\end{document}